\documentstyle[prb,aps,epsf,floats]{revtex}

\begin{document}
\renewcommand{\textfraction}{0.0}
\renewcommand{\floatpagefraction}{.7}
\setcounter{topnumber}{5}
\renewcommand{\topfraction}{1.0}
\setcounter{bottomnumber}{5}
\renewcommand{\bottomfraction}{1.0}
\setcounter{totalnumber}{5}
\setcounter{dbltopnumber}{2}
\renewcommand{\dbltopfraction}{0.9}
\renewcommand{\dblfloatpagefraction}{.7}

\draft

\twocolumn[\hsize\textwidth\columnwidth\hsize\csname@twocolumnfalse%
\endcsname

\title{Vortex lines in the three-dimensional XY model with random
  phase shifts}

\author{Mai Suan Li}
\address{Faculty of Engineering and Design, Kyoto Institute of Technology,
Matsugasaki, Sakyo-ku, Kyoto 606 Japan\\
Institute of Physics, 02-668 Warsaw, Poland}

\author{Thomas Nattermann}
\address{Institut f\"ur Theoretische Physik, Universit\"at zu K\"oln,
50937 K\"oln, Germany}

\author{Heiko Rieger}
\address{HLRZ c/o Forschungszentrum J\"ulich, 52425 J\"ulich, Germany}

\author{Moshe Schwartz}
\address{Raymond and Beverly Sackler Faculty of Exact Sciences, School 
of Physics and Astronomy, Tel Aviv University, Ramat Aviv, Tel Aviv 69978, 
Israel}

\date{\today}

\maketitle

\begin{abstract}
  The stability of the ordered phase of the three-dimensional XY-model
  with random phase shifts is studied by considering the roughening of
  a single stretched vortex line due to the disorder. It is shown that
  the vortex line may be described by a directed polymer Hamiltonian
  with an effective random potential that is long range correlated. A
  Flory argument estimates the roughness exponent to $\zeta=3/4$ and
  the energy fluctuation exponent to $\omega=1/2$, thus fulfilling the
  scaling relation $\omega=2\zeta-1$. The Schwartz-Edwards method as well
  as a numerical integration of the corresponding Burger's equation
  confirm this result. Since $\zeta<1$ the ordered phase of the
  original XY-model is stable.
\end{abstract}
]

\section{Introduction}

Systems with a continuous symmetry of the order parameter
are particularly susceptible to the influence of frozen-in disorder
\cite{dw}. In this paper we investigate the stability of the ordered
phase of the three-dimensional XY-model with random phase shifts with
respect to topological defects.

The Hamiltonian of this model  may be written
in the following form
\begin{equation}
{\cal H} = - J\sum_{\langle ij\rangle} 
\cos (\theta_i - \theta_j - A_{ij}) \;,
\label{eqn0}
\end{equation}
where $J$ is an effective coupling, the phase variables
$\theta_i\in[0,2\pi]$ are placed on a $3$--dimensional hypercubic
lattice and the sum runs over all nearest neighbor pairs $\langle
ij\rangle$. The variables $A_{ij}$ are quenched random phase shifts
(or random gauges) on the bonds connecting nearest neighbors. For
simplicity we assume, that the $A_{ij}$ on different bonds are
uncorrelated and gaussian distributed with mean zero and variance
${\sigma}$. Below it will be useful to go over to the continuum 
description in which $A_{ij}$ is replaced by the field  $\vec A (\vec r)$.

Model (\ref{eqn0}) describes XY-magnets with random Dzyaloshinskii-Moriya
interaction \cite{rsn}. Other realizations of this model are
3D-Josephson-junction arrays with positional disorder \cite{jja} and
vortex glasses \cite{mvg}. 

In the case of the so-called gauge glass model~\cite{mvg} one assumes
$A_{ij}$ to be uniformly distributed between $0$ and $2\pi$ but we
expect that our model with gaussian disorder is equivalent to the
gauge glass model when $\sigma$ is large enough.

In $d=2$ dimensions the model (\ref{eqn0}) shows a low temperature
weak-disorder phase with quasi long range order (QLRO) and, for
$\sigma=0$ a vortex-driven Kosterlitz-Thouless-like transition ~\cite
{kt,rsn,nskl} to the disordered phase. A finite amount of disorder
shifts the transition to lower temperatures and reduces the universal
jump in the spin-spin correlation exponent from $\eta=1/4$ for
$\sigma=0$ to $\eta=1/16$ at $\sigma_c=\pi/8$.
As it was shown recently ~\cite{nskl,noreentrant}, this transition is
not reentrant, contrary to earlier findings ~\cite{rsn}, but in
agreement with results of Nishimori for a model with slightly
different correlations of the random phase shifts ~\cite{ni}.

In $ d=3$ dimensions, it is easy to show that spin wave excitations,
which couple only to $\vec\nabla\cdot\vec A$, do not destroy true long
range at all temperatures and disorder strengths. To address the
question, whether this picture is qualitatively changed even for weak
disorder and low temperatures by allowing vortex configurations we
consider a {\it single} strechted vortex line in the presence of the
disorder field $\vec A$. Such a vortex line can be forced into the
system e.g. by appropriate boundary conditions. As long as the vortex
line remains self-affine with a roughness exponent $\zeta<1$, which
implies (for weak disorder and low T) a finite line tension, we
conclude, that the ordered phase remains stable. It turns out, that
this problem can be mapped approximatively onto a directed polymer
problem with long range correlated disorder. We find both from
analytical and numerical calculations, that the roughness exponent is
indeed about $0.75$, such that the ordered phase remains stable.
This is in agreement with results of Nishimori
~\cite{ni} for a slightly different probability weight for the
disorder.

\section{Spin waves and vortices}

In order to separate between spin wave and vortex degrees of freedom
we start from the continuum description of model (\ref{eqn0})
\begin{equation}
{\cal H}\; =\; \frac{J}{2}\; \int \, d^3 r \,  
| \vec{\nabla} \theta ( \vec{r} ) \, - \,\vec{A} ( \vec{r} )|^2 \;,
\label{eqn1}
\end{equation}
The quenched vector field $\vec{A} (\vec{r})$ is assumed to be Gaussian
distributed with zero average and correlations ($\alpha , \beta$ = 1, 2, 3)
\begin{equation}
\langle A_{\alpha} (\vec{r}) A_{\beta} (\vec{r'})\rangle \; = \; \sigma
\delta _{\alpha \beta} \delta (\vec{r} - \vec{r'}) \;,
\label{eqn2}
\end{equation}
where $\langle\cdots\rangle$ denotes the disorder average. 

The original model (\ref{eqn1}) is periodic in $\theta_i - \theta_j$
with the periodicity $2\pi$. In order to preserve this periodicity in
the continuum version (\ref{eqn2}) we have to allow for singularities
along which ${\bf \nabla}\theta$ jumps by $2\pi$. These singular
surfaces are bounded by vortex lines $l$ which are characterized by
their topological charge $m_l$ and their position vector
$\vec{R}_l(s)$.  It is convenient to decompose the $\vec\theta$-field into
a spin-wave part $\vec{\nabla}\theta_{sw}$ and a vortex part 
$\vec {\nabla}\theta _v$. 
The spin wave part is vortex free.  The vortex part of the vector field
${\bf \vec {\nabla}\theta_v}$ is defined by the saddle point equation
\begin{equation}
\vec{\nabla}\cdot(\vec{\nabla}\theta_v - \vec{A}(\vec{r})) = 0
\label{eqn3}
\end{equation}
and
\begin{equation}
\vec{\nabla} \times (\vec{\nabla} \theta_v) = 2\pi \vec{m}(\vec{r})\;,
\label{eqn4}
\end{equation}
where $\vec{m}(\vec{r})$ denotes the vortex density field which is non-zero
only along the singular lines $\vec{R}_l(S)$
\begin{equation}
\vec{m}(\vec{r}) = \sum_l m_l \int ds  \frac{d\vec{R}_l(s)}{ds}\delta^{(3)}
(\vec{r}-\vec{R}_l(s))
\label{eqn5}
\end{equation}
Eqs. (\ref{eqn4}) and (\ref{eqn5}) can be solved easily by introducing
a vectorpotential $\vec{a}(\vec{r})$, $\vec{\nabla} \times
\vec{a}(\vec{r}) = \vec{\nabla}\theta _v - \vec{A}(\vec{r})$. The
solution is
\begin{eqnarray}
\vec{\nabla}\theta _v(\vec{r}) = & 
\int d^3r'G(\vec{r}-\vec{r'})\{-\vec{\nabla '}
(\vec{\nabla '}\cdot\vec{A}(\vec{r'}))\nonumber\\
& + 2\pi \vec {\nabla '} \times \vec {m}(\vec{r'})\}
\label{eqn6}
\end{eqnarray}
with $\vec{\nabla}\cdot\vec{m}=0$. Here $\vec{\nabla'}$ denotes a
derivative with respect to the primed variable $\vec{r'}$, $\times$
represents a vector product.  The Green function $G (\vec{r})$
satisfies $\nabla ^2 G (\vec{r})=-\delta (\vec{r})$.  In three
dimensions $G (\vec{r})$ takes the form $G (\vec{r})=(4 \pi r)^{-1}$

Representing $\theta (\vec{r})$ as a sum of a vortex part $\theta_v
(\vec{r})$ and a spin-wave part $\theta _{sw} (\vec{r})$ one can show
that Hamiltonian (\ref{eqn1}) is decomposed into two independent spin
wave and vortex parts. In what follows we will be interested only in the
vortex part ${\cal H}_v$ which may be written in the form
\begin{eqnarray}
{\cal H}_{v } \; \; & = &\; \; {\cal H}_{vv} \; + \; {\cal H}_{vd} \;
\; , \nonumber\\
{\cal H}_{vv} \; \; & = & \; \; \frac{J \pi}{2} \int \int d^3 r d^3 r'
\frac{\vec{m}(\vec{r}) \cdot \vec{m}(\vec{r'})}{| \vec{r} - \vec{r'} |}
\;,\label{eqn7}\\
{\cal H}_{vd} \; \; &  = & \; \; \frac{J}{2} \; \int \int
d^3 r d^3 r' \frac{\vec{m}(\vec{r}) \cdot (\vec{\nabla'}\times\vec{A}(\vec{r'}))}
{| \vec{r} - \vec{r'}|} \;.\nonumber
\end{eqnarray}
The lattice version of such a vortex Hamiltonian for the gauge glass
has been derived recently \cite{by}.  Obviously, ${\cal H}_{vv}$ and
${\cal H}_{vd}$ correspond to the vortex-vortex and vortex-disorder
interactions respectively.  As in $d=2$ dimensions, the vortices
couple only to the $\vec{\nabla}\times
\vec{A}(\vec{r})=2\pi\vec{Q}(\vec{r})$, which can be considered as a
quenched random vector charge field with $\vec{\nabla}\cdot\vec{Q}=0$.

As follows from (3), $\langle\vec{Q}(\vec{r})\rangle =0$ and
\begin{equation}
\langle Q_{\alpha}(\vec{r})Q_\beta(\vec{r'})\rangle =\frac{\sigma}{4\pi}
[\partial_\alpha\partial_\beta - \delta_{\alpha \beta}
\nabla^2]\delta(\vec{r}-\vec{r'})
\end{equation}

Eqs.\ (8) and (9) would be the starting point for the statistical treatment
of model (1). In the partition function we had to integrate over all
possible configurations of vortex loops and vortex lines spanning 
the system. However, this task is much too difficult and remains to 
be done even for the pure system.

Instead, we follow here a much more modest approach and consider the
case of very strong dilution of vortex lines.  In order to test the
stability of the vortex free state with respect to vortex formation, it is
indeed sufficient to consider an isolated large vortex loop. Without
disorder, such a vortex loop costs an energy of the order $\sim\Sigma_{0} L\ln
L$ (see below) if $L$ is the radius of the loop $\Sigma_{0}\sim Jm^{2}$
denotes the bare line tension of vortex line. The configurational
entropy of the loop is also of the order $L\ln L$ and hence will
produce a negative free energy only at sufficiently high temperatures.
At these temperature the system is then disordered, the state of the
system is characterized by multiply entangled vortex lines and loops.

However, at low temperatures this mechanism does not work. A possible
source for the condensation of vortex loops here is the disorder. To
check this possibility, we consider in the next section a single
stretched vortex line which we allow to become rough under the
influene of the disorder. As long as the typical transverse distortion
$u\sim t^\zeta$ of a piece of length $t$ of the vortex line is
characterized by a roughness exponent $\zeta < 1$, the contribution
$\delta\Sigma_t$ of the disorder to the total vortex line tension,
$\Sigma$, from distortions on this length scale is of the order 
$-\Sigma_{0} u^2/t^2 \propto -t^{2(\zeta-1)}$ and hence small for large $t$.
  
Summation over the contributions $\delta\Sigma_t$ from all length
scales $b^{n} a$ ($a$ being the microscopic length scale or small 
scale lattice cut-off and $b$ the usual renormalization factor) 
between $n_{\rm min}=0$ and 
$n_{\rm  max}=\ln(L/a)/\ln b$ yields a finite value for 
$\delta\Sigma_{\rm total}$ which can be made arbitrarily small for 
decreasing disorder strength. Thus the line tension 
$\Sigma=\Sigma_0+\delta\Sigma_{\rm total}$ remains positive 
and the system is stable with respect to the condensation of vortex
lines. 

However, for $\zeta=1$ the disorder contributions to the line tension
are independent of $t$ and hence the energy per unit length will
vanish on a sufficiently large length scale $t_c$.
As a result, vortex loops of size $L > t_c$ will appear spontaneously and hence 
destroy the ordered phase. In this way we traced back the existence of a 
vortex free low temperature phase to the determination of the roughness 
exponent $\zeta$ of a single stretched vortex line.
We expect no difference in the case of a vortex loop as long as $t_c$ is
sufficiently large, which can be always acchieved for weak disorder.

\section{single vortex line Hamiltonion}

We consider in the following only a {\it single} stretched vortex line with
no overhangs, which means that we may set $\vec{R}(s) \to \vec{R}(t) =
(r_1(t), r_2(t), t)$, where ${\bf r}(t)=(r_1(t),r_2(t))$ describes the
transverse distortion of the vortex line from a straight configuration.
Then $\vec{m}(\vec{r})$ may be parametrized as
follows~\cite{kleinert}
\begin{eqnarray}
\vec{m}(\vec{r}) \; \; & = & \; \;  m \int dt
\frac{d\vec{R}}{dt} \delta ^{(3)} ( \vec{r} - \vec{R} (t)) \; \; ,
\nonumber\\
\frac{d\vec{R}}{dt} \; \; & = & \; \; \Bigl( \; \frac{\partial r_{1}}
{\partial t} , \; \frac{\partial r_{2}}{\partial t}, \; 1 \Bigr) \; \; ,
\label{eqn8}
\end{eqnarray}
${\cal H}_{vv}$ decribes now the elastic self-interaction of the vortex 
line ${\cal H}_{vv} \to {\cal H}_{el}$.
\begin{eqnarray}
{\cal H}_{el} \; \; & = & \; \; \frac{J \pi}{2} \;  m^2
\int \int \frac{d\vec{R}(t)d\vec{R}(t')}{|\vec{R}(t)-R(t')|}
\nonumber\\
d\vec{R}(t)d\vec{R}(t')
& = & \; \; \Bigl( \; \frac{\partial {\bf r}}
{\partial t} . \frac{\partial {\bf r}}{\partial t'} \; + \; 1
\; \Bigr) \, dt dt' \; .
\label{eqn9}
\end{eqnarray}

Expanding ${\cal H}_{vv}$ in Eq.(\ref{eqn7}) up to quadratic terms in
${\bf r}$ and $\partial {\bf r}/\partial t$ and omitting an irrelevant
constant corresponding to the energy of straight lines one has
\begin{eqnarray}
{\cal H}_{el} \; =  \; \frac{J \pi}{2} \,
m^2\, \int \int & & \frac{d t d t'}{| t - t' |} \, [ \,
\frac{\partial {\bf r}}{\partial t} \cdot \frac{\partial {\bf
    r}}{\partial t'}
\nonumber \\
& &  - \, \frac{({\bf r}(t) - {\bf r}(t'))^2}{2 | t - t' |^2} \, ] +
O(r^4)\; \; .
\label{eqn12}
\end{eqnarray}
It is convenient to go over to the Fourier transform
${\bf r} (t)=\int ^l_0e^{i\omega t}{\bf r}_\omega\frac{d\omega}{(2\pi)}$
\begin{eqnarray}
{\cal H}_{el} \; \; & = & \; \; \frac{Jm^2}{2}
\int \frac{d\omega}{(2\pi)} \omega^2 f (a\omega) {\bf r}_\omega . {\bf r}_{-\omega} \;,
\label{eqn13}
\end{eqnarray}
where $a$ denotes a small scale lattice cut-off and
\begin{equation}
f(\epsilon) =
\int^{\infty}_{\epsilon}\frac{dx}{x}[(1+\frac{1}{x^2})\cos x
-\frac{1}{x^2}]\approx -\ln \epsilon - \frac {3}{4}
\label{fepsilon}
\end{equation}
for $\epsilon \to 0$.

Next we consider the correlations of the vortex disorder interaction. 
For this purpose we rewrite ${\cal H}_{vd}$ as
\begin{equation}
{\cal H}_{vd} = \int d \vec {R}\cdot \vec{V}(\vec{R})\;,
\label{eqn14}
\end{equation}
where
\begin{equation}
\vec{V}(\vec{R}) = J\pi m \int d^3r\frac{\vec{Q}(\vec{r})}{|\vec{R}-\vec{r}|}
\label{eqn15}
\end{equation}
Clearly $\langle \vec {V}\rangle =0$ and 
\begin{eqnarray}
\langle V_{\alpha} (\vec {R})V_\beta(\vec {R'})\rangle = 
& & \sigma \frac {\pi J^2 m^2}{2}\frac{1}
{|\vec {R}-\vec {R'}|}  \{\delta_{\alpha\beta} + \nonumber \\ 
& + & \frac{(R_\alpha - R'_\alpha)
(R_\beta - R'_\beta)}{(\vec {R} -\vec {R'})^2}\}\label{eqn16}
\end{eqnarray}
Since the correlator (\ref{eqn16}) always appears under the integral
over $\vec{R}$, it is convenient to rewrite the right hand side
(\ref{eqn16}) as

\[\frac{\pi}{2}\sigma{(Jm)}^2(\frac{\partial^2}{\partial R_\alpha \partial R'
_\beta}|\vec{R}-\vec{R'}|+\frac{\delta_{\alpha\beta}}{|\vec{R}-\vec{R'}|})
\]

The first contribution in this expression leads to terms in the correlator of 
${\cal H}_{vd}$ 
which vanish by assuming periodic boundary conditions. Thus we omit it in
the following. For small gradients
$|\frac{\partial r}{\partial t}| \ll 1$, ${\cal H}_{vd}$
can finally be rewritten as

\begin{equation}
{\cal H}_{vd} = \int dt V({\bf r}(t), t)
\end{equation}
with
\begin{equation}
\langle V({\bf r},t)V({\bf r}',t')\rangle = \sigma (Jm)^2|({\bf r}-{\bf r}')^2
+(t-t')^2|^{-1/2}
\label{correlatorintime}
\end{equation}

Thus, the random potential $V(\vec {R})$ which interacts with the vortex 
line is long range correlated.

\section{Flory arguments and beyond}

In the following we want to estimate the conditions under which the
random potential can destabilize the vortex line.  We begin with a
{\it straight} line of length $t$, which has an elastic energy $E_{el} =
\frac{J\pi}{2} m^2 t \ln t/a$ as follows easily from (\ref{eqn9}). The
typical fluctuations of the disorder energy follows from (\ref{eqn16})
as $E_{dis}=\pm mJ\sqrt {\pi \sigma}(t\ln t/a)^{1/2}$. Since rare
fluctuations increase $E_{dis}$ only by a factor $(\ln t)^{(1/2)}$,
the system is always stable with respect to the formation of straight
vortex lines.

Next we allow for displacement ${\bf r}(t)$ from the straight
configuration such that $|\frac{\partial {\bf r}}{\partial t}|\ll 1$. If
on scale $t$ the vortex line is self-affine with $u=\langle {\bf r}^2\rangle 
^{1/2}
\sim t^\zeta$ with $\zeta < 1$, we can use a Flory-argument to
estimate $\zeta$.  The elastic energy is then increased by $\Delta
E_{el}\approx \frac{Jm^2\pi}{2}t^{-1}u^2\ln t/a$ which has to be
compared with $E_{dis}$. This yields
\begin{equation}
u\sim [\sigma t^3 (\ln t/a)^{-1}]^{1/4}\;,
\label{eqn17}
\end{equation}
i.e. $\zeta=3/4$. Although the Flory argument is rather crude, it
seems to be safe enough to conclude $\zeta < 1$, i.e. for weak
disorder the vortex line is self-affine and hence stable (only $\zeta
= 1$ would signal an instability with respect to vortex generation).

To go beyond the Flory argument one usually maps these types of directed 
polymer problems to Burger's equation with (correlated) noise.
However, in the present case this is not possible in the strict sense because 
of the long
range elastic self-interaction of the vortex line, which leads to the
factor $f(a\omega) \approx \ln \frac{1}{a\omega}$
in (\ref{fepsilon}). Since this corresponds to a logarithmic increase of the
stiffness of the vortex line with the length scale, we neglect the
$\omega $-dependence of this factor in the following completely and set 
equal to 
$f_0$. 
If then the roughness exponent $\zeta$ is less than one, it will be less
than one by including $f(a\omega)$. In fact, it is even safe to
assume that the value of $\zeta$ is unchanged by replacing
$f(a\omega)$ by $f_0$ since for small $a\omega$ $f(a\omega)$ diverges
only logaritmically.

With this simplification the free energy $h({\bf r},t)$
of a vortex line of length $t$ and ${\bf r}={\bf r}(t)-{\bf r}(0)$
obeys the Burger's equation with noise
\begin{equation}
\frac{\partial}{\partial t}h({\bf r},t)=\nu{\bf \nabla} ^2 h+
\frac{\lambda}{2}
({\bf \nabla} h)^2 + V({\bf r},t)
\label{burgers}
\end{equation}

where $\nu={T}/{2Jm^2f_0}$ and $\lambda={1}/{Jm^2f_0}$.

Besides of the roughness exponent $\zeta$ there is a second exponent
$\omega$, which describes the sample to sample fluctuations 
$\langle h^2({\bf r},t)\rangle -\langle h({\bf r},t)\rangle^2\sim t^{2\omega}$
of the free energy. Since the correlator (\ref{correlatorintime})
is non-local in time, the validity of the scaling relation
$\omega=2\zeta-1$ is not guaranteed \cite{medina}.

If we give $h({\bf r},t)$ another interpretation, namely that of the
height profile of a growing interface in a co-moving frame, eq.\
(\ref{burgers}) is known as the KPZ-equation \cite{kpz}. The roughness
exponent $\alpha$, the growth exponent $\beta$ and the dynamical
exponent $z$ of this {\it surface} are related to the directed polymer
exponents by
\begin{eqnarray}
\zeta \; \;& =& \; \; \beta/\alpha \; \; = \; \; 1/z \; \; \; , \nonumber\\
\omega \; \; &=&\; \; \beta \; \; = \; \; \alpha/z\; \; \;.
\label{scalingrel}
\end{eqnarray}

We have first attempted to use the standard one-loop renormalization
group treatment for the Burger's equation~\cite{medina}.
Unfortunately, this approach does not yield (as happens also for the
case of uncorrelated noise in three dimensions) stable fixed points
from which one could calculate critical exponents. We study the
Burger's equation with the correlated noise (eq.21) therefore here by
a method that proved useful for the same problem with uncorrelated
noise ~\cite{ss}.

Our first step here is to 
describe our system as a system of dynamical variables affected by
noise that is uncorrelated in time. This is necessary in order
to go over from the Langevin-like Burger's equation to the
Fokker Planck equation, needed in the method mentioned above.

This step is achieved by considering the $V$ to be also dynamical variables
coupled to a noise $\xi$ that is uncorrelated in space and time.

In Fourier components we write eq.\ (\ref{burgers}) in the form
\begin{equation}
\frac{\partial h_{\bf q}}{\partial t} = -\nu q^2h_{\bf q}-\frac{\lambda}{2\sqrt 
{\Omega}}
\sum_{\bf l} {\bf l}\cdot
({\bf q}-{\bf l})h_{\bf l}h_{{\bf q}-{\bf l}}-V_{\bf q}(t)
\end{equation}
where $\Omega$ is the transverse area of the system(to be taken eventually
to infinity) and add to it an equation for $V_q(t)$
\begin{equation}
\frac{\partial V_{\bf q}}{\partial t} = -|{\bf q}|V_{\bf q} + \xi _{\bf q}(t)
\end{equation}
where $\langle\xi_{\bf q}(t)\rangle =0$ and 
$\langle\xi_{\bf q}(t){\bf \xi}_{\bf -q}(t')\rangle =
2D\delta(t-t')$ and 2$D$ is 
short for $4\pi\sigma(Jm^2)$. It is easily verified that $V_{\bf q}(t)$ has
the requiered correlations given by eq.(19).

We write now the Fokker-Planck equation for the joint probability density 
to have a given $h$ and $V$
configuration, $P\{h_{\bf q}, V_{\bf q}\}$
\begin{eqnarray}
\frac{\partial P}{\partial t}
=\sum_{\bf q}\Biggl\{\frac{\partial}{\partial V_{\bf q}}
\biggl(D\frac{\partial}{\partial V_{\bf q}}
+|{\bf q}| V_{\bf q}\biggr)
+\frac{\partial}{\partial h_{\bf q}}\biggl(\nu q^2h_{\bf q}\nonumber\\
+\frac{\lambda}{2\sqrt{\Omega}}\sum _{\bf l}{\bf l}
\cdot({\bf q}-{\bf l})h_{\bf l}
h_{\bf q-l}
+V_{\bf q}\biggr)\Biggr\}P
\end{eqnarray}

We will be interested in the steady state averages:
$\langle h_{\bf q}h_{-{\bf q}}\rangle_s$, 
$\langle h_{\bf q}V_{-{\bf q}}\rangle_s$, and $\langle V_{\bf q}V_
{\bf -q}\rangle_s$ 
and in some 
characteristic 
frequencies to be defined later. We assume now that the exact values of 
$\langle h_{\bf q}h_{-{\bf q}}\rangle_s,
\langle h_{\bf q}V_{-{\bf q}}\rangle_s$ and 
$\langle V_{\bf q}V_{-{\bf q}}\rangle_s$ are known and 
given by $X_{\bf q}^{-1},\Delta
_{\bf q}^{-1}$ and $\Lambda _{\bf q}^{-1}$. In order to calculate such 
quantities, 
we need some 
form of a perturbation expansion in which the Fokker-Planck operator $O$, 
acting on $P$
on the right hand side of eq.(24), will be broken into two parts: one part 
$O_0$ 
that is simple enough 
and another part $O-O_0$, that is small enough. We may expect that if we 
choose $O_0$ in such a way, that it already gives the exact result for 
the three steady state averages, the corrections to those quantities
in perturbation theory, are going to be small giving sense to the whole
expansion.

Our choise for $O_0$ is

\begin{eqnarray}
O_0 = \sum _{\bf q} \omega_{\bf q}^{(1)}\frac{\partial}{\partial V_{\bf q}}
[\Lambda_{\bf q}^{-1}
\frac{\partial}
{\partial V_{-{\bf q}}}+V_{\bf q}]\nonumber\\
+\omega_{\bf q}^{(2)}\frac{\partial}{\partial h_{\bf q}}[X_{\bf q}^{-1}
\frac{\partial}{\partial h_{-{\bf q}}}+h_{\bf q}]\nonumber\\
+[\omega^{(1)}_{\bf q} + \omega_{\bf q} ^{(2)}]
[\Delta _{\bf q}^{-1}\frac{\partial}{\partial h_{\bf q}}
\frac{\partial}{\partial V_{-{\bf q}}}]
\end{eqnarray}

It is easily verified, that regardless of the choice of the $\omega '_{\bf q}
s$, 
$O_0$ produces $\langle h_{\bf q}h_{-{\bf q}}\rangle_s=X_{\bf q}^{-1},
\langle h_{\bf q}V_{-{\bf q}}\rangle_s
=\Delta _{\bf q}^{-1}$
and $\langle V_{\bf q}V_{-{\bf q}}\rangle_s=\Lambda_{\bf q}^{-1}$.

A direct inspection of eq. (24), shows that the form of the $V$
dependent part of $O$ is identical to the form of the correspoding part of 
$O_0$. Therefore we must identify
$\Lambda_{\bf q}^{-1}=\frac{D}{|{\bf q}|}$ and $\omega^{(1)}_{\bf q}=|{\bf q}|$.

Next we calculate in perturbation theory 
$\langle h_{\bf q}h_{-{\bf q}}\rangle_s$ and 
$\langle h_{\bf q}V_{\bf -q}\rangle_s$. We find 
\begin{equation}
\begin{array}{rcl}
\langle h_{\bf q}h_{-{\bf q}}\rangle_s &=& X^{-1}_{\bf q} + C_1\{X^{-1}_{\bf l},
                             \Delta^{-1}_{\bf l},\omega^{(2)}_{\bf l}\} \\
\langle h_{\bf q}V_{-{\bf q}}\rangle_s & = & 
                           \Delta^{-1}_{\bf q}+C_2\{X^{-1}_{\bf l},
                           \Delta_{\bf l}^{-1},\omega_{\bf l}^{(2)}\} \;.
\end{array}
\end{equation}
Since $X_{\bf q}^{-1}$ and $\Delta _{\bf q}^{-1}$
are assumed to be exact, we get now, for a given set $\{\omega_{\bf q}^{(2)}\}$,
two equations $C_1=C_2=0$ expressing the fact that the perturbation expansion 
does not change the zero order value. Actually it could be expected
that for any choice of positive $\omega_{\bf q}^{(2)'}s$, the above equations 
for the 
$X'_{\bf q}s$ and $\Delta '_{\bf q}s (C_1=C_2=0)$, would give the exact 
result. It is clear however, that even for statics, we may use the choice of
the $\omega^{(2)'}_{\bf q}s$ in order to control the smallness of the additional 
terms and thus we will be able to obtain our $X'_{\bf q}s$ and $\Delta '
_{\bf q}s $ from 
{\it low} order perturbation. Assuming that such a choice is possible we write
$C_1$ and $C_2$ to second order to obtain the two equations:
\begin{equation}
\begin{array}{rcl}
0=&-&[\omega^{(1)}_{\bf q}+\omega^{(2)}_{\bf q}]\Delta^{-1}_{\bf q}
     +\frac{[\nu q^2-\omega^{(2)}_{\bf q}]^2}
           {[\omega^{(1)}_{\bf q}+\omega^{(2)}_{\bf q}]}
      \Delta_{\bf q}^{-1}\\
  &+&\frac{[\nu q^2-\omega^{(2)}_{\bf q}]D}{|{\bf q}|[\omega^{(1)}_{\bf q}
          +\omega^{(2)}_{\bf q}]} - \frac{D}{|{\bf q}|}\\
  &+&\frac{\lambda^2}{{(2\pi)}^2}
     \left\{\int d^2{\bf l}\frac{{\bf l}\cdot({\bf q}-{\bf l})}
        {[\omega^{(1)}_{\bf q}+\omega^{(2)}_{\bf q-l}
       +\omega_{\bf l}^{(2)}]}{\bf q}
     \cdot(-{\bf l})X_{\bf l}^{-1}\right\}\Delta_{\bf q}^{-1}
\end{array}
\label{longeq1}
\end{equation}
and
\begin{equation}
\begin{array}{rcl}
0=&-&[2\nu q^2-\frac{{(\nu q^2)}^2}{\omega^{(2)}_{\bf q}}]X^{-1}_{\bf q}\\
  &+&[\nu q^2-\omega^{(2)}_{\bf q}]
     [\frac{1}{\omega_{\bf q}^{(2)}}
      +\frac{1}{\omega^{(1)}_{\bf q}+\omega_{\bf q}^{(2)}}]
     \Delta ^{-1}_{\bf q}\\
  &+&\frac{D}{|{\bf q}|[\omega^{(1)}_{\bf q}+\omega^{(2)}_{\bf q}]}\\
  &+&\frac{{\lambda}^2}{2{(2\pi)}^2}
     \int d^2l\frac{{\bf l}\cdot({\bf q}-{\bf l})}
      {[\omega_{\bf l}^{(2)}+\omega^{(2)}_{{\bf q}-{\bf l}}
       +\omega^{(2)}_{\bf q}]}
      {\bf l}\cdot({\bf q}-{\bf l})X_{\bf l}^{-1}X^{-1}_{{\bf q}-{\bf l}}\\
  &+& 2{\bf q}\cdot({\bf l}-{\bf q})X^{-1}_{{\bf q}-{\bf l}}X_{\bf q}^{-1}
\end{array}
\label{longeq2}
\end{equation}

We fix now $\omega_{\bf q}^{(2)}$ to be the characteristic frequency associated with
the correlations of $h_{\bf q}$ at different times
\begin{equation}
\omega^{(2)}_{\bf q}=\frac{\langle h_{\bf q}h_{-{\bf q}}\rangle_s}{\int_0^\infty\langle h_{-\bf q}
(0)h_{\bf q}(t)\rangle dt}
\end{equation}

where $\langle A(0)B(t)\rangle $ is defined for $t>0$ to be
\begin{equation}
\begin{array}{rl}
\langle A(0)B(t)\rangle =
\int\,& A\{h'_{\bf q},V'_{\bf q}\}
      P_s\{h'_{\bf q},V'_{\bf q}\}
      P\{h'_{\bf q},V'_{\bf q};h_{\bf q},V_{\bf q};t\}\\
      & B\{h_{\bf q},V_{\bf q}\} Dh'_{\bf q} DV'_{\bf q}Dh_{\bf q}DV_{\bf q}\;.
\end{array}
\end{equation}
$P_s$ is the static (steady state) distribution and $P\{h'_{\bf q}, V'_{\bf q};
h_{\bf q},V_{\bf q},t\}$
is the solution of the Foker-Planck equation for initial conditions of absolute
certainty that the value of $h_{\bf q}$ is $h'_{\bf q}$ and $V_{\bf q}$ is 
$V'_{\bf q} $
($\Pi_{\bf q}\delta(h_{\bf q}-h'_{\bf q})\delta(V_{\bf q}-V'_{\bf q})$ 
is the initial distribution).
Our lowest order Fokker-Planck operator, $O_0$, is thus 
chosen also to give the correct characteristic frequency. We calculate
now the characteristic frequency to second order and demand again that
the correction to the zero order term is zero. This yields
\begin{equation}
\begin{array}{rcl}
0&=&[2-\frac{\nu {\bf q}^2}{\omega_{\bf q}^{(2)}}][\omega^{(2)}_{\bf q}-\nu 
    {\bf q}^2]X_{\bf q}^{-1}\\
 &+&\frac{\lambda ^2}{2(2\pi)^2}
    \int d^2l\frac{{\bf l}\cdot({\bf q}-{\bf l})}
    {[\omega^{(2)}_{\bf l}+\omega^{(2)}_{{\bf q}-{\bf l}}]}
    2{\bf q}\cdot({\bf l}-{\bf q})X^{-1}_{\bf l}X^{-1}_{\bf q}\\
 &+&\frac{\lambda ^2}{2(2\pi)^2} \int  d^2l \frac{\omega ^{(2)}_{\bf q}
    {\bf l}\cdot({\bf q-l})}{[\omega ^{(2)}_{\bf l}
        +\omega ^{(2)}_{\bf q-l}][\omega_{\bf l}^{(2)}
        +\omega ^{(2)}_{\bf q-l}
        +\omega _{\bf q}^{(2)}]}\\
   &&\quad\cdot[2{\bf q}\cdot({\bf l}-{\bf q})X_{\bf l}^{-1}X_{\bf q}^{-1}
     +{\bf l}\cdot({\bf q}-{\bf l})X_{\bf l}^{-1}X_{\bf q-l}^{-1}]\\
   &-&\frac{\omega _{\bf q}^{(2)}\Delta^{-1}_{\bf q}}{\omega_{\bf q}^{(1)}}
\end{array}
\label{longeq3}
\end{equation}

Apart from the last term, eq.\ (\ref{longeq3}) is identical to the
corresponding equation in \cite{ss}.

We assume now that for small $q$
\begin{equation}
\begin{array}{rcl}
\omega^{(2)}_{\bf q} & \propto & q^{z}      + {\rm corrections}\\
X_{\bf q}            & \propto & q^{\Gamma_1} + {\rm corrections}\\
\Delta_{\bf q}       & \propto & q^{\Gamma_2} + {\rm corrections}
\end{array}
\end{equation}
The exponents $z$ and $\Gamma_1$ are related to the exponents
$\zeta$ and $\omega$ by the relations $\zeta=1/z$ and
$\omega=(\Gamma_1-2)/2z$.  We solve equations (\ref{longeq1}),
(\ref{longeq2}), and (\ref{longeq3}) to leading order and obtain the
folowing results. The scaling relation
\begin{equation}
z=\frac{6-\Gamma_1}{2}
\end{equation}
is obeyed .(The last term in eq.(\ref{longeq3}) is negligible compared
to the dominant terms for small q). This is the familiar scaling
relation
\begin{equation}
\omega=2\zeta-1
\label{scaling0}
\end{equation}
Eqs.(\ref{longeq1}) and (\ref{longeq2}) allow now for a simple power
counting solution (In contrast to ref.13 where a power
counting solution was impossible).

We obtain 
\begin{equation}
\Gamma_1=\frac{10}{3}, \quad z=\frac{4}{3} , \ \Gamma_2=2
\end{equation}

It is interesting to note that the result for $\Gamma_2$ is not
affected by $\lambda$ being not equal to zero and this is because the
last term in eq.(\ref{longeq1}) is negligible, for small $q's$,
comopared to the dominant terms. Consideration of higher order terms
in the expansion, (in particular, the most dominant $q$ dependence in
each order) seems to suggest that this results will still hold to any
order of perturbation theory. In terms of the more familiar directed
polymer exponents we find
\begin{equation}
\zeta = \frac{3}{4} \quad \omega=\frac{1}{2}
\end{equation}
so that the Flory result still holds.

\section{Numerical results}

In this section we will obtain the roughness exponent for our 
model by integrating  the
(2+1)-dimensional KPZ equation (\ref{burgers}) numerically. 
 The simplest quantity to investigate
is the surface width
\begin{equation}
W = \langle\overline{h^2 } - \overline{h}^2\rangle^{1/2}\;,
\end{equation}
where the bar and angular brackets $\langle\ldots\rangle$ denote a
spatial and noise averages respectively.  One expects for $W_L$
the following scaling form~\cite{fv}
\begin{equation}
W \; \; = \; \; L^{\alpha} \, f(t/L^z) \; \; ,
\label{wscale}
\end{equation}
where $f(x) \, \sim \, x^{\beta}$ with $\beta = \alpha /z$ for $x\ll1$
and $f(x) \, \sim \, const.$ for $x\gg1$; $\beta$, $\alpha$ and $z$
are the growth, roughness and dynamic exponent of the interface
respectively, c.f.\ (\ref{scalingrel})

In analogy to (\ref{wscale}) the height-height correlation function 
\begin{equation}
C_L(\vec{r},t) = 
\langle[h (\vec{r}+\vec{r'}, t+t' ) - h (\vec{r'},t')]^2\rangle^{1/2}
\label{corrf}
\end{equation}
should scale as
\begin{equation}
C_L(r,t) \; = \; r^{\alpha} \tilde{g}(r/t^{1/z},L/t^{1/z})\;.
\label{corrscale}
\end{equation}

We do not take the scaling relation
\begin{equation}
\alpha \; + \; z \; \; = \; \; 2 \; \;, \label{scaling}
\end{equation}
which is equivalent to  (\ref{scaling0}),
for granted in our investigation. Thus, to obtain a full set of the
critical exponents we have to calculate at least two of them.

Different theoretical approaches~\cite{medina,zhang,henf} lead to
different dependences of the critical exponents on $\rho$ in
(1+1)-dimensions. To check theoretical predictions the numerical
simulations have been carried out for the KPZ equation~\cite{peng},
the ballistic deposition~\cite{mj1,peng,amar}, the directed
polymer~\cite{peng} and the solid-on-solid model~\cite{amar}. There is
still a controversy between the simulation results. For example, the
results of Amar {\em et al.}~\cite{amar} obtained for the ballistic
deposition and restricted solid-on-solid models agree with the
prediction of Medina {\em et al.}~\cite{medina} but conflict with the
prediction of Zhang~\cite{zhang}. Numerical studies~\cite{peng} of the
effect of the long-range spatially correlated noise on the KPZ
equation and the related directed polymer problem, on the other hand,
give a good agreement with the prediction of Hentschel and
Family~\cite{henf}. Temporally correlated noise in the absence of
spatial noise correlations has also been investigated
numerically~\cite{temporal} for the (1+1)-dimensional case.

Theoretical results for (2+1)-dimensional models with correlated noise
are still lacking. Meakin and Jullien~\cite{mj1,mj2} introduced a
hopping model of ballistic deposition, in which particles were
deposited on the growing surface following a Levy flight distribution
such that the distance $x$ (along the surface) from the previous site
was calculated as $x = r^{-1/f}$ where $r$ is a random number between
zero and 1. Equating the exponent $f$ to $2$ as in (\ref{correlatorintime})
led to results~\cite{mj1} for the exponents $\alpha$ and $\beta$ which
were roughly in agreement with the prediction of Medina {\em et al.}
for (1+1) dimensions.  A weak dependence of these exponents on $f$ has
been found in (2+1)-dimensional model~\cite{mj2}.

In the simulations of Meakin and Jullien the link between the
deposition process and the noise was not obvious. Thus the scaling
behavior in (2+1)-dimensional interface growth and the directed
polymer model with spatially correlated noise is at the present time
unclear.  We consider this problem solving the KPZ equation
numerically. The spatial derivatives in (\ref{burgers}) are discretized
using standard forward-backward differences on a hypercubic grid with
a lattice constant $\Delta x$.  Eq. (\ref{burgers}) is integrated by the
Euler algorithm with time increments $\Delta t$. Denoting the grid
points by $\vec{n}$ and the basic vectors characterizing the surface
by $\vec{e}_1, \ldots , \vec{e}_d$ we arrive at the discretized
equation~\cite{moser}
\begin{eqnarray}
& \tilde{h} (\vec{n},\tilde{t}+\delta\tilde{t}) =
\tilde{h} (\vec{n},\tilde{t}) \nonumber\\
& + \frac{\textstyle\Delta\tilde{t}}{\textstyle\Delta\tilde{x}^2} 
\sum_{i=1}^{d} \; \{ \, [ \tilde{h} (\vec{n}+\vec{e}_i,\tilde{t}) -
2 \tilde{h}(\vec{n},\tilde{t}) +
\tilde{h}(\vec{n}-\vec{e}_i,\tilde{t})] 
\nonumber\\
& + \frac{\textstyle1}{\textstyle8} 
[ \tilde{h} (\vec{n}+\vec{e}_i, \tilde{t}) \, - \,
\tilde{h} (\vec{n}-\vec{e}_i, \tilde{t}) ]^2 \, \} \; + \;
\sqrt{3 \Delta \tilde{t}} \, \eta (\vec{n}, \tilde{t}) \; \nonumber\\
\label{discrete}
\end{eqnarray}
Here one uses dimensionless quantities $\tilde{h} = h/h_0, \tilde{x} = x/x_0$
and $\tilde{t} = t/t_0$ where the natural units are given by
\begin{equation}
h_0 = \frac{\nu}{\lambda} , \; \; t_0 = \frac{\nu ^2}{\sigma ^2 \lambda ^2} 
\, , \; \; x_0 = \sqrt{\frac{\nu ^3}{\sigma ^2 \lambda ^2}} \, , \;
\; \sigma ^2 = 2D/\Delta x^d \; \; .
\end{equation}
Similar to $V(\vec{r},t)$ in the Fourier space the 
renormalized noise $\eta (\vec{n}, \tilde{t})$ has 
the correlation
\begin{equation}
\langle  \eta (\vec{k}, \omega) \eta (\vec{k'}, \omega') \rangle  \; 
\; \sim \; \; \frac{\delta (\vec{k}+\vec{k'}) \delta (\omega + \omega')}{k^2
+ \omega ^2} \; \; .\label{noisecorr}
\end{equation}
To create the spatially correlated noise we follow 
Peng {\em et al.} \cite{peng}. We first generate a standard white
(or Gaussian uncorrelated) noise $\eta _0 (\vec{n}, \tilde{t})$, then
carry out the Fourier transformation for spatial and temporal variables
to obtain $\eta _0(\vec{k},\omega )$. We define
\begin{equation}
\eta (\vec{q},\omega ) \; \; = \; \; ( k^2 + \omega ^2 )^{-1/2} 
\eta _0 (\vec{k}, \omega) \; \; .
\end{equation}
The noise $\eta (\vec{n},\tilde{t})$ is obtained by Fourier transforming
$\eta (\vec{k},\omega )$ back into the real space. It is easy to
check that $\eta (\vec{k},\omega )$ obained by this way satisfies
(\ref{noisecorr}).

\begin{figure}[hbt]
\epsfxsize=\columnwidth\epsfbox{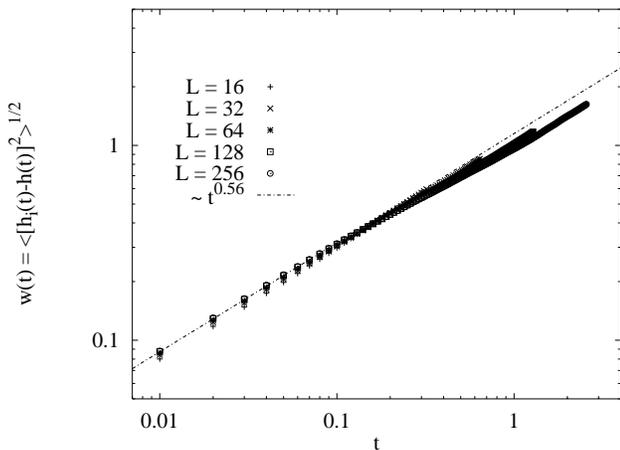}
\caption{Log-log plot of the width 
  $W_L(t)$ vs $t$ for various system sizes $L$.  For smaller times the
  data are roughly independent of system size with a slight tendency
  to a smaller slope for increasing $L$. For larger times (i.e.\ when
  $t\sim L^z$) significant finite size effects appear (i.e.\ the
  scaling function $\protect{\tilde{w}}$ in (\protect{\ref{scalew}})
  is not constant any more. Therefore we put more weight on the data
  points at smaller times in order to determine $\beta$, which yields
  $\beta=0.56\pm0.02$ shown as dotted line.}
\end{figure}

In our simulations we chose $\Delta x$, $\nu$ and $\sigma$ to be the
same as in Ref. \cite{moser}, namely $\Delta x = 1, \nu =0.5$ and
$\sigma =0.1$. To be sure that we are in a strong-coupling regime we
chose $\lambda =\sqrt{600}$. The data we show in the following are
averaged over 10000 samples (i.e.\ different noise realizations) for
the smallest system size ($L=16$) to 50 samples for the largest system
size ($L=256$).

First we determine the exponent $\beta$ and the dynamical exponent $z$
from simulations of large system sizes on short time scales, i.e.\ we
focus on a regime where the correlation length $\xi$ is still small
compared to the system size $L$:
\begin{equation}
\xi\propto  t^{1/z}\ll L.
\end{equation}
In this regime finite size effects should be negligible, which we had
to check explicitely by analyzing different system sizes since we do
not know $z$ a priori. For the time dependence of the roughness we
expect the scaling form (\ref{wscale}) to hold, which can be cast into
the form
\begin{equation}
W_L(t) = t^\beta \tilde{w}(t/L^z),\label{scalew}
\end{equation}
with $\tilde{w}(x)=const.$ for $x\to0$ and $\tilde{w}(x)\sim x^{-1}$
for $x\to\infty$. For short times $t\ll L^z$ one therfore expects 
$W_L(t)\propto t^\beta$. Fig. 1 shows the time dependence of the width
$W_L(t)$ for various system size $L$ at short time scales. We find
\begin{equation}
\beta = 0.56 \pm 0.02\;. \label{beta}
\end{equation}
This value is much larger than $\beta \approx 0.240$ for the
three-dimensional KPZ equation with the white noise \cite{moser}, as
we expect it to be since correlated noise always increases $\beta$.

In order to obtain the dynamical exponent $z$ from the short time
behavior we note that for $t\ll L^z$ the height-height correlation
function $C(r,t)$ (\ref{corrf}) should have the scaling form
\begin{equation}
C(r,t) \; = \; r^{\beta z} g(r/t^{1/z}) \; ,
\end{equation}
where we denoted 
$\lim_{(tL^{-z})\to0}\tilde{g}(r/t^{1/z},L/t^{1/z})=g(r/t^{1/z})$ 
and used $\alpha=\beta z$.

\begin{figure}[hbt]
\epsfxsize=\columnwidth\epsfbox{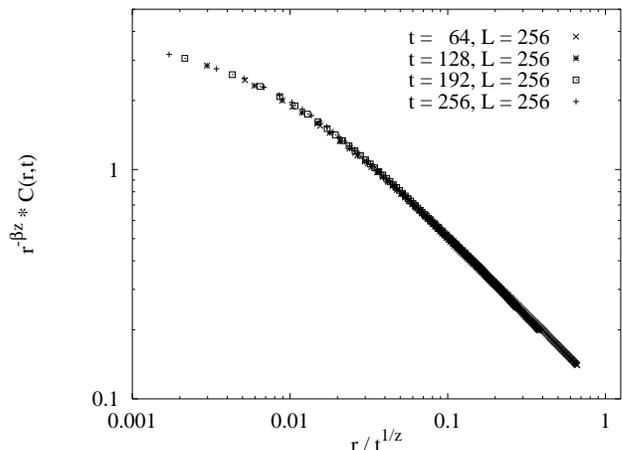}
\caption{Scaling plot for the height-height correlation C(r,t)
  function. We fixed $\beta$ to be $0.56$, as determined before
  (\protect{\ref{beta}}) and varied $z$ in order to achieve the best
  data collaps, which yiels $z = 1.25 \pm 0.05$.}
\end{figure}

If we take $\beta=0.56$ as determined before we are left with one
fitting parameter by which we should achieve a data collapse when
plotting $r^{-\beta z}C(r,t)$ versus $r/t^{1/z}$. In fig.\ 2 we show
the result of this procedure, which yields
\begin{equation}
z=1.25\pm0.05\;.\label{z}
\end{equation}

\begin{figure}[hbt]
\epsfxsize=\columnwidth\epsfbox{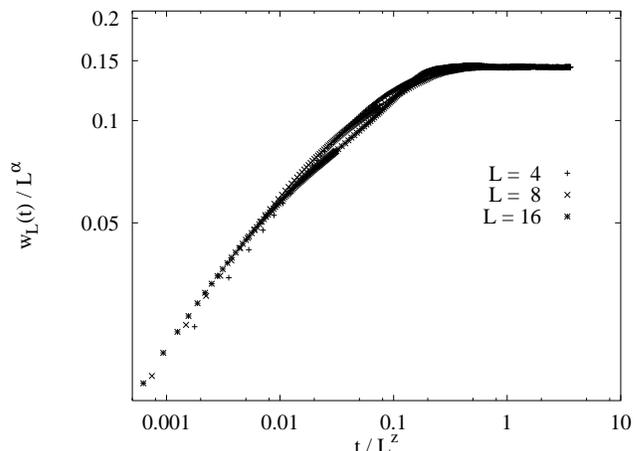}
\caption{Scaling plot of $W_L(t)$ according to the scaling form
  (\protect{\ref{scalew}}). The choice $\alpha=0.71\pm0.01$ and
  $z=1.25\pm0.03$ yields the best data collaps, which is shown.}
\end{figure}

Up to now we have restricted ourself to a regime, where finite size
effects are negligible. In order to get an independent estimate for
the exponents reported above we performed also a simulation for much
longer times and small enough system size, so that the correlation
length becomes indeed comparable to the system size (i.e.\ $t\sim
L^z$). Using the finite size scaling form (\ref{wscale}) we determine the
exponents $\alpha$ and $z$.  In fig.\ 3 we show a scaling plot of
$W_L(t)$ which yields
\begin{equation}
\alpha=0.71\pm0.01
\end{equation}
and $z=1.25\pm0.03$, from which one concludes
$\beta=\alpha/z=0.56\pm0.03$. These values agree very well with the
values reported in (\ref{beta}) and (\ref{z}) for the short time
simulations.

Using scaling relation (\ref{scalingrel}) we have 
\begin{equation}
\omega = 0.56 \pm 0.02\qquad{\rm and}\qquad\zeta = 0.78 \pm 0.04
\label{polexp}
\end{equation}
for the vortex lines in our model.  The value of $\zeta$ is larger
than the corresponding value $\zeta \approx 0.62$ for the directed
polymer paths with uncorrelated noise in three dimensions. Our
numerical result is compatible within the error bars with the
crude estimate $\zeta=3/4$ obtained by the Flory argument and by the
more elaborate calculation in the last section. The numerical value of
$\omega$ is slightly larger than the calculated $\omega=1/2$.

Note that within the error bars relation (\ref{scaling}) is still
valid even in the presence of the temporal correlation of noise.  This
relation may hold true only for some subset of all possible
correlators (\ref{noisecorr}). At present we have to leave this
question open.

\section{Summary and Conclusion}

In the present paper we have considered the stability of the ordered
phase of the XY-model with random phase shifts. After decomposing the
Hamiltonian into a spin-wave and a vortex part, we have considered in
particular the roughening of a single stretched vortex line due to the
disorder. It turned out that the effective random potential acting on
the vortex line is long range correlated. Using a Flory argument and
the Schwartz-Edwards-method, we have determined the roughness exponent
$\zeta=3/4$ and the energy fluctuation exponent $\omega=1/2$, which
fulfill the scaling relation $\omega=2\zeta-1$.  These findings have
been confirmed by integrating numerically the Burger's equation.
Since $\zeta<1$, there will be no spontaneous condensation of vortices
for weak disorder and hence we conclude, that the ordered phase is
stable.

A further interesting application of our result is that on the
XY-model in a random field \cite{rfield,gh}. From
the results of Villain and Fernandez \cite{vf} for this model {\it
  without} vortices one finds $\sigma(t)\approx\sigma^* t^{d-2}$,
where $\sigma^* = O(\frac{4-d}{d-2})$. Using this $t$-dependence of
$\sigma$ in the Flory-argument one concludes $\zeta=1$ for $2<d<4$,
apart from a logarithmic correction. From this one should expect that
one is in a marginal situation, which deserves further investigation
\cite{knh}.

\acknowledgments 

This work has been performed within the Sonderforschungsbereich SFB
341 K\"oln--Aachen--J\"ulich supported by the Deutsche
Forschungsgemeinschaft DFG. 
M.S.L acknowledges financial support from the SFB 341, the Japan
Society for Promotion of Sciences and the Polish agency KBN
(grant number 2P302 127 07).
T.~N.\ and M.~S.\ have been supported in part by the German Israeli
Foundation (GIF).

\end{document}